# Time and its arrow: an empiricist's perspective

Stephen Boughn⃰

Departments of Physics and Astronomy, Haverford College, Haverford, PA 19041

The nature of time has beguiled philosophers for nearly three millennia. There are myriad types of time including cosmological time, geological time, biological time, physiological time, psychological time, physical time, historical time, and even theological time. I'm sure there are others. My brief essay concerns time in physics and barely scratches the surface of this fascinating subject. Even so, I hope that the pragmatism of an experimental physicist might help provide a perspective that is often absent in treatises by contemporary philosophers and physicists. This is especially the case for the notion of the *arrow of time*. The paradox of the arrow of time arises because the laws of physics are invariably time reversal invariant, in which case we are led to ask what determines the direction of time. Investigations of time's arrow are usually framed in the context of a mathematical formalism where the parameter $t$ represents time, and then proceed to logical analyses of how the direction of time emerges. The implicit assumption of these deliberations is that time is an ontological property of nature. On the other hand, for an empiricist like me, it is absolutely crucial to frame any such discussion in terms of direct human experience rather than on a parameter in a mathematical model of nature. This, in short, is the motivation for my essay. I'll begin with a folk-historical account of the concept of time simply for the purpose of providing context for my subsequent remarks and conclude with an argument that the paradox of the arrow of time is an artificial problem that needs no resolution.

**A Story of Time**

It is not difficult to imagine how our ancient ancestors arrived at the notion of time. Soon after humans acquired the ability to count, they began counting days, sunrises and sunsets, and perhaps recorded them with marks on a piece of wood. This number count probably was the first measure of time and might have led to a notion of cyclic

---

⃰ sboughn@haverford.edu

2time. However the days evolved, for example, the sun didn't rise and set at the same locations on the horizon yet the same sequence of locations repeated itself the following year, a longer temporal cycle. Another measure of time was the position of the sun as it moved across the sky. Such a quantity was surely useful for predicting the coming of darkness, a skill that must have been essential for survival. But time was not just cyclic. From year to year, humans aged and eventually died. On longer time scales ancient societies died out, lakes dried up, and game animals disappeared. So it seems that the notion of time for our early ancestors was quite likely a combination of secular (time's arrow) and periodic (time's cycle) phenomena.[1] From these examples it is clear that the measure of time is directly related to change. Things happen and time is way of ordering these happenings.

Another much discussed aspect of time is the notion of past, present, and future. While this is more the purview of philosophers than of physicists, I'll also offer a folk-account of this facet of time, if for no other reason than, to indicate why it is not particularly relevant for *physical time*. Events we experience are often imprinted in our memories that provide records of the events. The present is coincident with the creation of memories, the past is identified by the totality of our memories, and the future corresponds to often anticipated events for which there are, as yet, no memories. Records are certainly not limited to human memories. Most physical events produce records, for example, the geological and fossil records that we find on the surface of the earth, the astronomical records of long ago supernovae explosions in the form of the radiation and matter created in these events, and historical records written in countless volumes. Our sense is that we live in the present rather than in the past or the future, as if we are in a flowing river of time, where upstream constitutes the future and downstream, the past. I suppose this sort of metaphor gives us the impression of the flow of time. In any case, I consider the apparent flow of time to be more a subject for psychology and neuroscience than for physics. Even so, physicists are not totally isolated from this motif. We are well aware of what we are currently observing, what observations we have already made, and what observations we plan for the future.

---

[1] Stephen Jay Gould (1987) gave a fascinating account of the discovery of geological time in a book entitled *Time's Arrow, Time's Cycle* that included the importance of theology to the principals involved.

**Time's Arrow**

While it seems unlikely that our primitive ancestors thought about the *direction of time*, they were aware that the sun and moon always rise in the east and set in the west and might have wondered whether this would ever change. They certainly observed that children grew up and eventually aged and died but it's doubtful that this observation led to a question about the direction of time. The same might also be said about the observation that an egg breaks when dropped on a stone or the hard ground. It is unlikely that they wondered why broken eggs never reassemble themselves and hop back into one's hand. They simply knew that the former does happen and the latter never happens.

Philosophers as far back as Plato, and I'm sure before, pondered the nature of time. However, I'm restricting my discussion to time's arrow in the context of contemporary physics and will therefore skip the deliberations of philosophers from the classical Greek period through the Renaissance up to the epoch of modern science beginning with Newton. Classical (Newtonian) and modern (quantum) physics mark the introduction of mathematical models of nature that usually take the form of equations containing a parameter *t* denoting time. It then became feasible to address the notion of the arrow of time with a rigor made possible by mathematical laws of nature.[2] This brings us to the main theme of my essay, physics and the arrow of time.

**Physical Time**

From the above "story of time" it is apparent that early measures of time consisted of counting the number of days, months, and years from the observed solar and lunar cycles. As our ancestors advanced, shorter intervals of time were determined from a variety of simple devices including sundials, water clocks, candle/incense clocks, hour glasses, simple pendula, and even human heartbeats. The evolution of clocks from simple devices to much more elaborate instruments continues to this day. My point is that what a physicist means by time comes directly from how time is measured.[3] Einstein emphasized this in the context of his 1905 special theory of relativity. He realized that the only way Maxwell's theory of electrodynamics could be made self-consistent was to reject the notion of universal time and instead insist that time is precisely the quantity measured by clocks.  In so doing, he was able to make sense of the fact that the speed of light is a constant independent of the velocity of the source or observer. Einstein's ideal device was a light clock that is elegant yet simple and easily grasped.  A premise of this essay is that physical time is precisely the quantity that clocks measure, no more no less, and so discussions relating to time, certainly relating to the arrow of time, should be framed in terms of the observations of clocks and not solely in terms of mathematical models of nature.

---

[2] However it's important to keep in mind that these "laws" are, in fact, models, i.e., human inventions.

[3] Today, a second of time is *defined* as the duration of 9,192,631,770 periods of the an electromagnetic wave emitted via the transition between the two hyperfine levels of the unperturbed ground state of a caesium-133 atom.

**Physical Time's Arrow**

Early in their education, physics students learn that Newton's laws of mechanics are time reversal invariant.  This means that if one replaces $t$ with $-t$ in, for example, Newton's second law, the result is precisely the same second law (if the force is also time reversal invariant, which is so for all fundamental forces).  The apparent implication is that if a certain process follows Newton's laws, then that process in reverse still follows Newton's laws.  Now let's see if we can frame the discussion in terms of direct experience.  Imagine making a movie of a series of events that presumably conform to Newtonian mechanics and then run the movie in reverse.  If nature is time reversal invariant, this backward sequence should also satisfy Newton's laws.  In terms of human experience, we immediately see a problem. Virtually any movie run backward displays a bizarre series of events that never occurs.  Recall the case of a broken egg spontaneously reassembling itself and hopping from the ground into a person's hand.  Why doesn't this ever happen?  Why does time run in one direction and not in the other?  This, in short, is the *paradox of the arrow of time* but notice that the source of the paradox is the time reversal invariance of a mathematical model, in this case Newton's laws.

Let's first consider the much simpler case of the Moon revolving west to east around the Earth with a period of about one month. If one plays a movie of this scenario in reverse, the moon would simply revolve around the Earth from east to west.  While this vision conflicts with our experience, it doesn't jar our sensibilities.  The reverse case certainly satisfies Newton's time reversal invariant laws and all that it would take to realize this situation would be to reverse the direction of the moon's initial velocity $v$ even though it's hard to imagine circumstances that would bring this change about. Reversing the moon's velocity is not time reversal invariant since $v(-t) = -v(t)$. Indeed, many have argued that initial conditions determine the direction of time. On the other hand all arrows of time, human experience, the geological record, biological evolution, etc., seem to point in the same direction so what initial conditions are we talking about? Some maintain that the relevant initial conditions are those imposed in the very early universe at the time of the Big Bang while others maintain that the expansion of the universe after the Big Bang determines the arrow of time[4]; although, it's not immediately obvious how such conditions could affect the arrow of time in relatively isolated locales in the universe.

More appropriate examples arise from classical thermodynamics and statistical mechanics. Suppose we were to place a hot block of material in contact with a cold block. After some time (as measured by a clock) both blocks will reach the same warm temperature. This scene played in reverse would have two warm blocks placed in contact with one becoming cold and the other becoming hot, an occurrence that would certainly jar the senses. As another example, consider an open container of perfume that is placed in a closed room. After an extended period of time, the perfume will have evaporated with the scent pervading the entire room. In the reverse process an empty container is placed in a scented room. After the same interval of time, the container will be filled with perfume in a room with no aroma, another jarring affair. Neither of these events ever occurs. Why not? Ludwig Boltzmann explained, via statistical mechanics, that both of these reverse sequence of events are consistent with time reversal invariant Newtonian dynamics and could happen if the initial conditions could be precisely arranged. However, this is virtually impossible to do and, otherwise, there is a vanishingly small probability that such conditions would ever occur spontaneously. This argument is directly related to the increase in entropy as required by the second law of thermodynamics and many have convincingly argued that the arrow of time is, therefore, *determined* by the second law. (We'll return to this topic later.)

Another example of a time invariant "law of nature" is Maxwellian electrodynamics. Like Newton's second law, Maxwell's equations are time reversal invariant. A simple but perplexing case involves an oscillating charged particle. One solution of Maxwell's equations for this system is an outgoing, periodic electromagnetic wave that travels at the speed of light to great distances. In fact, this is what happens when radio waves are broadcast from an antenna. However, there is another solution to Maxwell's equation with $t \to -t$ that corresponds to an outgoing wave traveling backward in time to an infinite distance at $t = -\infty$. This bizarre solution can also be given an equally bizarre interpretation of an incoming electromagnetic wave from the infinite past at infinite distance that arrives at the charge the instant the oscillation begins. Maybe the latter interpretation removes the notion of traveling backward in time but it certainly stands causality on its head. Some might argue that the time reversed solution can be eliminated by an appeal to initial conditions but it's hard to wrap one's mind around prescribing, for this simple system, initial conditions that are set at an infinite distance in the infinite past. The usual practice is simply to ignore this solution, which seems reasonable to me.

---

[4] See, for example, Thomas Gold (1962) and Julian Barbour (2020).

Now let's get back to the spontaneous reassembly of a broken egg. Eggs are complicated systems consisting of atoms and atomic interactions; therefore, we need a model of atoms to proceed.[5] This leads us to quantum mechanics, a fundamental theory that is nominally considered to be time reversal invariant.[6] In particular, the Schrödinger equation is time reversal invariant in the sense that if $t \to -t$ then the complex conjugate of the Schrödinger wave function $\psi$ satisfies the time reversed Schrödinger equation. Because observables only depend on the absolute value of $\psi$, the Schrödinger equation is effectively time reversal invariant. Does this mean that the time reverse of an egg breaking also satisfies the Schrödinger equation? Not really. In the case of the reversed lunar orbit an initial condition that reverses the velocity is all that's required to reverse the sequence of observations of the moon. The quantum mechanical wave function is something entirely different. It's not a physical property of a system but rather is assigned to a system in order to make probabilistic predictions of a subsequent observation of that system.[7] The sequence of observed events of a broken egg spontaneously reassembling itself cannot be described by quantum mechanics. According to Freeman Dyson (2016), "The wave function only describes probabilities for the result of an observation in the future. When the observation moves into the past, the result is described by facts instead of probabilities." Quantum mechanics is simply incapable of describing the sequence of events in a movie run in reverse nor even of the sequence of events of the original movie for that matter. Therefore, whether or not the Schrödinger equation is time reversal invariant is irrelevant.

---

[5] Of course, the previous examples also involve atoms but the systems are simple enough that ignoring their quantum properties is not a problem.
[6] To be sure, there are aspects of quantum theory, e.g., weak interactions, that violate time reversal invariance; however, these violations are relatively small and few suggest that they are responsible for what humans perceive as the direction of time.

It's possible to muse about time's arrow simply in terms of what we observe but the *paradox* of the arrow of time arises directly from the time reversal invariance of our mathematical models of nature. So let me offer a general comment about the concept of time in these models. To be sure a temporal parameter $t$ appears in the classical and quantum laws of physics but what is the relation of this $t$ to time (other than *time* begins with the letter $t$)? Does $t$ capture all that we attribute to time? I would argue no. Rather $t$, in some sense, should be regarded as one of many parameters in a set of mathematical equations that physicists have created to model our observations of the physical world. Why then do we think of $t$ as representing time? This brings us back to clocks. Remember clocks are invariably cyclic mechanical or electrodynamic devices, among the sorts of systems that the mathematical models of classical mechanics and electrodynamics were created to describe. The same can be said about secular sequences of events. Therefore, it's not surprising that the $t$ parameter in those models is proportional to the temporal periods of clocks and the rate of secular change. For this reason it's usually fine, for all practical purposes, to identify $t$ with time. On occasion, however, this can lead one astray. The path from time reversal invariance of physical models to the paradox of time's arrow is one of these. In the case of the Schrödinger equation, the identification of $t$ with time contributes to the notion that quantum

---

[7] How these assignments are made is not prescribed by quantum mechanics. What physicists do in practice is to build up a catalog of transcriptions from operationally prepared systems to wave functions based on past experience. Only then is one able to assign a wave function to a particular system, which can then be used to make statistical predictions of subsequent observations. See H. P. Stapp's "The Copenhagen Interpretation" for an excellent account of Bohr's pragmatic interpretation of quantum mechanics. [Stapp 1972]

mechanics is fully capable of describing a reverse sequence of observed events, which I have argued is not the case.

**Entropy and Time**

Perhaps the most common explanation of the arrow of time invokes entropy and the second law of thermodynamics. In fact, many claim that the second law resolves the paradox of time's arrow for all systems including the case of the spontaneous reassembly of a broken egg. To be sure, there are many detailed arguments in this vein, too numerous to mention [8]. So, as before, I'll resort to a folk-account to provide context for my remarks.

According to the second law, the entropy of a closed system always increases until it arrives at a state of thermodynamic equilibrium where entropy is the highest. I'm sympathetic with the notion that our perception of the arrow of time is related to the second law. Boltzmann's explanation of the impossibility of the time reverse scenario of a perfume bottle in a closed room is certainly convincing. On the other hand, humans are examples of systems whose entropy often decreases and we certainly don't experience our arrow of time as reversing with respect to other systems around us. Of course humans aren't closed systems but interact with others and one must take the entropy of these other systems into account. Taken to the extreme, the entropy of the entire universe might be required to determine the *true* arrow of time and might also explain why we perceive the universe as expanding and not contracting.[9] (This harkens back to the initial conditions at the beginning of the universe.) Certainly the second law and the notion of entropy have proved useful in dealing with a wide range of physical phenomena, including the physics of black holes.[10] According to the second law, the change in entropy $S$ of a closed system cannot decrease, i.e., $\Delta S \geq 0$ and since *time* is a measure of the ordering of *changes*, this might also be expressed as $\Delta S/\Delta t \geq 0$. So the second law is certainly related to time but it seems implausible to me that an inequality provides a basis for interpreting time. The second law simply implies that entropy $S$, like time $t$, is intrinsically related to change.

---

[8] See, for example, Craig Callender in *Stanford Encyclopedia of Philosophy* (2016).
[9] In his provocative new book, *The Janus Point: A New Theory of Time,* Julian Barbour gives a fascinating account of these issues. In fact, it was Barbour's book that led me to write this essay.
[10] However, one should keep in mind that the second law and entropy are models we created in our quest for understanding nature and not necessarily fundamental properties of nature herself.

Einstein famously extolled the importance of thermodynamics and maintained it was foremost in his mind as he developed relativity theory. In his autobiographical notes he opined (Einstein 1949):

> A theory is the more impressive the greater the simplicity of its premises is, the more different kinds of things it relates, and the more extended is its area of applicability. Therefore the deep impression which classical thermodynamics made upon me. It is the only physical theory of universal content concerning which I am convinced that, within the framework of the applicability of its basic concepts, it will never be overthrown.

When I first encountered Einstein's view of thermodynamics, I was mystified. This youthful skepticism was heightened when I learned that thermodynamics was simply a coarse grained model of a more fundamental theory, statistical mechanics. If so, why did Einstein consider it to be a theory that "will never be overthrown". Then, a few years ago, I read a fascinating account of entropy by Elliott Lieb and Jakob Yngvason. In a lengthy 1999 paper in *Physics Reports,* they demonstrated that the second law follows from quite general properties of physical systems.

Lieb and Yngvason consider pairs of system states $X$ and $Y$ such that $Y$ is "adiabatically accessible"[11] from $X$. They then demonstrate that "a list of all possible pairs" of such states "can be simply encoded in an entropy function $S$ on the set of all states of all systems (including compound systems), so that when $X$ and $Y$ are related at all, then $Y$ is accessible from $X$ if and only if $S(X) \leq S(Y)$." This definition of entropy is independent of any underlying statistical model and even whether or not the systems are composed of interacting atoms. No assumptions about the arrow of time and course graining are necessary nor are concepts like heat and temperature. The only requirements are a, presumably empirical, list of accessible states and a set of logical conditions on accessibility. The second law follows naturally from this construction of the entropy function.[12]

---

[11] In this context, "adiabatic processes do not have to be very gentle" but rather are simply "defined by processes whose only net effect on the surroundings is exchange of energy with a mechanical source." [Lieb and Yngvason 1999]

[12] This is less than a bare bones account of their analysis presented in a 96 page document. A much shorter paper intended for a more general physics audience appeared in *Physics Today* in 2000.

The generality of the Lieb and Yngvason analysis helped me understand the wide applicability of thermodynamics and the second law, as Einstein long ago noted. It also provides a lucid explanation of the meaning of entropy. Entropy is quite simply a way of encoding change in physical systems and the second law follows from empirical observations of change. Time, as I've characterized it, is simply a way of enumerating empirical change so it's no surprise that time and entropy are related. They are both ways of expressing what we observe. There is no need to go further. Characterizing time in terms of entropy or entropy in terms of time does not enhance our empirical understanding of these concepts. The "paradox" of the arrow of time is a statement about the mathematical properties of our models of nature. There is no paradox in what we observe.

**Final Remarks**

I began this essay by insisting that discussions of time should be framed in terms of direct experience. So rather than pondering the direction of the arrow of time, one might simply ask why a certain reversed sequence of events never occurs. There is a straightforward answer to the question of why we don't observe the perfume molecules in an aroma filled room condensing back into an open container. Boltzmann's argument that, given Newton's laws, there is a vanishing probability that such a scenario would ever occur spontaneously provides a convincing answer. Furthermore, the entropy of the system would have to decrease in order for it to happen, a clear violation of the second law of thermodynamics. Therefore, it seems reasonable to claim that the second law prevents the reverse scenario from occuring. Notice, however, there is no mention of the arrow of time. What about the case of the spontaneous reassembly of a broken egg? Why does this never occur? I've argued that the language of quantum mechanics is incapable of directly addressing this question; however, the concept of entropy is still applicable and again one might answer that it could never happen without violating the second law. Even though I have no idea how to calculate the total entropy change in this case, I won't dispute that some clever theorist could do so and find that a decrease of total entropy is required. The Lieb/Yngvason construction of entropy from a empirical list of accessible states renders this entirely plausible but, again, the notion of the arrow of time needn't be mentioned.

While I find such arguments convincing, I'm somewhat uncomfortable about using physics to address questions that ask *why* something occurs rather than something else. For example, why we don't see herds of unicorns roaming the plains of Nebraska? One might answer that physics doesn't preclude such an occurrence but rather it's simply happenstance that nature does not realize it. Both unicorns and broken egg reassembly are very complicated systems but the possibilities of their occurrences seem to be governed by different considerations. Yet, for the myriad of even more complex systems, is there a distinct dividing line between "a happenstance of nature" and "a violation of the $2^{nd}$ law"? Perhaps, but I tend to doubt it. Lieb and Yngvason's construction of an entropy function depends on a list of accessible states determined by what we observe. We physicists usually take the world as we experience it and then create models to describe what we observe. Finally, I suppose that we could express the broken egg question without using "why". For example, "Is the spontaneous reassembly of an egg inconsistent with the laws of physics?" On the other hand, if this is a reasonable question then so should the more general question: "Is quantum mechanics inconsistent with what we don't observe?", a question that raises my level of discomfort considerably.

My distress with this sort of question aside, I still claim that the arrow of time has no place in such a discussion. In fact, if we avoid the insistence that the parameter $t$ in our mathematical models is identical with all we mean by time, then the paradox of the arrow of time evaporates. It is common practice to refer to our hallowed mathematical models as *laws of nature* and so it's no wonder that time is often taken to be a distinct component of the *real world*. This strikes me as a step backward toward Isaac Newton's much decried absolute time that Einstein replaced with operationally defined clocks. Also, our fundamental physical theories are more than just mathematical models. In addition to the aforementioned initial conditions, our models all have accompanying interpretive statements that inform us how to use them. These statements are usually expressed in common language that is wholly outside the theoretical formalism. Consider, for example, the operational statements governing the use of the Schrödinger wave function to predict the outcomes of future experiments. The notion of time reversal invariance makes little sense for such interpretive statements, yet another reason why our physical theories of nature are incapable of answering questions about the arrow of time.

Don't get me wrong; investigations of the mathematical structure of fundamental theories are an essential aspect of physics and certainly those relating to $t$ are of crucial importance. For example, in classical and quantum mechanics there is a fundamental relation between time $t$ and energy $E$ while in quantum field theory the consequences of charge, parity, and time reversal symmetry ($CPT$) are critically important. Also, the origin of the $t$ parameter in our theories is certainly intimately related to the time measured by clocks and it's fine to identify $t$ with time for all practical purposes. However, the strict identification of $t$ with all of what we understand as *time* leads directly to the paradox of the arrow of time and it is with this identification that I find fault. Time is not a monolithic entity. In order to encompass the many sorts of time mentioned above, it is necessary to embrace a pluralistic view. I suspect that the failure to recognize this pluralism is responsible, in part, for the protracted discussions about *the* real meaning of time and its arrow.

It's obvious that my pragmatic stance as an experimental physicist might strike many as the sort of radical empiricism that William James espoused.[13] I freely admit that this is the case but hope that my somewhat novel and, perhaps, provocative approach provides a different perspective that may be of some interest.

## Acknowledgements

I thank my two muses, Freeman Dyson and Marcel Reginatto, for encouraging my pursuit of understanding the foundations of quantum mechanics and many friends, especially Juan Uson and Jeff Kuhn, who have tolerated this pursuit. Thanks to Elliot Lieb and especially Marcel Reginatto for commenting on an earlier version of the paper and to Steve Weed for planting the seed for this essay in my mind.

---

[13] I recently expressed my views in a much longer essay entitled "On the Philosophical Foundations of Physics: An Experimentalist's Perspective" (2019).